# Genes Preferring Non-AUG Start Codons in Bacteria

Anne Gvozdjak[1,2] and Manoj P. Samanta[2,3,4]

1. Bellevue High School, 10416 SE Wolverine Way, Bellevue, WA 98004
2. Coding for Life Science, Redmond, WA 98053
3. Systemix Institute, Redmond, WA 98053
4. samanta@homolog.us

## Abstract

Here we investigate translational regulation in bacteria by analyzing the distribution of start codons in fully assembled genomes. We report 36 genes (infC, rpoC, rnpA, etc.) showing a preference for non-AUG start codons in evolutionarily diverse phyla ("non-AUG genes"). Most of the non-AUG genes are functionally associated with translation, transcription or replication. In *E. coli*, the percentage of essential genes among these 36 is significantly higher than among all genes. Furthermore, the functional distribution of these genes suggests that non-AUG start codons may be used to reduce gene expression during starvation conditions, possibly through translational autoregulation or IF3-mediated regulation.

## Introduction

Understanding the regulation of gene expression at the transcriptional and translational levels is a key step in deciphering the information contained by genomes. Whereas inexpensive, high-throughput technologies (ChIP-seq, RNAseq) are helping in extensive experimental investigation of transcriptional regulation [1,2], the measurement of translational regulation using high-throughput mass spectroscopy [3] is less widely adopted. Bioinformatics provides another avenue to leverage the rapidly falling cost of DNA sequencing to explore translational regulation. Prior to 2000, computational researchers identified a number of relevant patterns through the comparative analysis of gene sequences [4,5] . With the availability of ~250,000 prokaryotic genomes, thousands of eukaryotic genomes, and additional metagenomic sequences, those patterns can now be studied at a significantly larger scale, and new patterns may also be identified.

Among the three phases of translation, namely initiation, elongation, and termination, initiation is the rate-limiting step [6,7]. In prokaryotes, the Shine-Dalgarno sequence upstream of the start codon was considered to be the primary regulatory mechanism for initiation [8,9], but whole-genome sequencing changed this understanding. Leaderless genes, once considered rare [9], appear to be far more abundant [10] and are argued to provide a more ancestral initiation procedure [11]. Although the exact initiation mechanism in these genes is not understood, the nucleotides around the start codon likely play important roles.

The above realization brought renewed attention to start codons. Whole genome sequencing observed a widespread presence of GUG, UUG, and other non-AUG start codons in bacteria [12–15]. Irrespective of the start codon sequence, the same methionine-containing initiator tRNA always binds to it. Therefore, the differences in start codons may be used to regulate the rate of translation. Empirical studies have indeed shown that the rate of gene expression varies greatly depending on the start codon being used [16]. For instance, the well-studied gene infC uses an unconventional start codon in many evolutionarily diverse bacterial genomes to autoregulate its translation [17–20].

In this work, we looked for other genes showing a preference for non-AUG start codons by analyzing the distribution of start codons in fully assembled bacterial genomes from diverse phyla. Our analysis identified 36 genes (infC, rpoC, rnpA, etc.) with a preference for non-AUG start codons. Functionally, they are often associated with translation, transcription and replication. In *E. coli*, this group of genes carries a significantly larger proportion of essential genes than across all *E. coli* genes. Furthermore, based on data from existing literature, we argue that non-AUG start codons in these genes may help their regulation at the translational level, possibly as a response to nutrient stress.

## Results

This analysis started with 210,130 publicly available prokaryotic genomes, from which all genomes with incomplete assemblies or annotations were excluded. From the remaining 11,904 genomes, the start codon nucleotide sequences of 3,818,377 gene occurrences were extracted. This raw data produced a preliminary distribution of start codon frequencies across bacterial phyla (Figure 1). Overall, about 85% of gene occurrences use AUG as their primary start codon, followed by GUG (10%) and UUG (< 5%). Actinobacteria has a substantially lower percentage of AUG start codons, in agreement with the prior studies [13,21].

To investigate possible evolutionary patterns among various genes' start codon frequencies, this analysis was further restricted to the genomes from 30 genuses covering various large bacterial phyla (Table 1). Restricting the analysis to genuses with high total gene occurrences reduced potential noise from annotation errors [21]. The set of chosen genuses included 10 gammaproteobacteria and firmicutes each, and 10 from the other remaining phyla. For each gene within a genus, a normalized value ("non-AUG ratio") was introduced, measuring the ratio of its non-AUG start codon occurrences to its total occurrences within the genus. Genes with a higher "non-AUG ratio" across numerous genuses were considered more desirable. Therefore, a minimum cutoff of a 0.5 "non-AUG ratio" was implemented here.

Four genes (infC, rpoC, rnpA, hisS) were above this cutoff in at least half of the 30 genuses. It is encouraging to find infC at the top of this list. Not only is this gene involved in translation initiation itself, but its unconventional start codon has been the subject of extensive research [17–20]. Very little systematic research regarding the start codons has been conducted on the other three genes.

A total of 36 genes were above the cutoff in at least a third of the genuses (Table 2). They are referred to as "non-AUG genes" in the following text. To compare the phylogenetic distribution of start codon preferences, the percentage presence of various start codons for these genes in different genuses is displayed in Figure 2.

A number of interesting patterns emerge from Table 2 and Figure 2. The translation initiation factor infC maintains a non-AUG start codon in most genuses, but it is also joined by another initiation factor (infB) in a subset of the genuses. A similar pattern is observed among RNA polymerase components rpoC and rpoB.

Additionally, a large fraction of the genes are involved in translation, but they appear to play roles in only a subset of pathways. The genes rpsT, rpsN, and rpsC are all components of the 30S ribosomal subunit, whereas rlmH, rsmH, and rsmG are all methyltransferases targeting 16S rRNA, a component of the same subunit. Moreover, rimP is involved in the maturation of the same subunit. In contrast, no component of the 50S ribosomal subunit showed up, apart from typA assisting in its assembly at the low temperatures. The other translational genes in Table 2 are translation elongation factor tuf and ribosomal silencing genes rsfS, hflX, and rsgA.

Several other genes are involved in various aspects of transcription, translation and DNA replication. This set includes rnpA cleaving precursor tRNAs as part of the RNase P ribonuclease complex; hisS and argS ligating tRNAs with amino acids; grpE regulating protein folding; secF and secD assisting in protein export; pnp helping in mRNA degradation; nusB involved in transcription antitermination; dnaA taking part in chromosomal replication; and radA, ubrC, and recJ repairing DNA replication errors.

The remaining eight of the 36 non-AUG genes are not related to cellular information storage and processing pathways (i.e. translation, transcription, replication). They include atpF as an ATP synthase component; plsX involved in lipid metabolism; murE taking part in cell wall biogenesis; ispG performing isoprenoid biosynthesis; crcB reducing fluoride toxicity; pxpB catalyzing the cleavage of 5-oxoproline to L-glutamate; carA hydrolyzing glutamine to glutamine; and fdhD acting as a sulphur carrier necessary for formate dehydrogenation. Overall very few genes from the entire metabolic pathway was present in this set of 36 genes.

Among the non-AUG genes reported here, 47% (17 out of 36) are essential in *E. coli*, whereas deletion of 9 others lead to slowed growth. In comparison, only 4% of all *E. coli* genes are essential.

The non-AUG genes identified here yielded multiple members previously researched for their translational autoregulation. Indeed, genes such as rpsT, coding for the 30S ribosomal protein S20, and infC, coding for the bacterial translation initial factor IF3, have both been specifically investigated for the correlation between their use of a non-standard start codon and their translational autoregulation using negative feedback loops [17,22,23]. In both cases, replacing the start codon with AUG resulted in derepression and an increase in gene expression. The other genes known for autoregulation or IF3-mediated suppression of expression are rpoB, rpoC, dnaA, uvrC, pnp, and recJ [24–27]. In the case of rnpA, experimental evidence suggested potential translational autoregulation, although this was not further investigated [28].

## Discussions

Early gene sequencing in bacteria showed the presence of non-AUG start codons in about 8-9% of all genes. This estimate increased substantially to 20-40% after complete bacterial genomes became available [12,13,21]. Researchers in both pre- and post-genomic eras attempted to understand the regulatory roles of these alternate start codons [16–18,29]. Here we discuss our results in the context of those early and recent reports.

A simple model of the translational machinery would argue that the alternate start codons are there to reduce the expression levels of the corresponding genes. It is easy to see how the tRNA-codon binding would weaken for the non-canonical start codons, given that the same methionine-containing initiator tRNA binds to all start codons. Empirical studies confirm that the rate of gene expression reduces with the use of non-AUG start codons [16]. Based on this model, the genes identified in this work should have a general need for reduced expression in diverse phyla, thus explaining their preference for non-AUG start codons in general.

However, this explanation has two inconsistencies [30]. First, in isolated examples of ribosomal proteins (e.g. rpsM), changing the start codon from GUG to AUG reduced translation by seven-fold. Second, the genes involved in translation are among the most abundant in a cell. Therefore, it does not appear that non-AUG start codons are used to maintain a general reduction of their expression. Below we offer a refined model that takes this functional difference of the non-AUG genes into account.

Based on observations of the preservation of many of its common components in all domains of life, translation appears to have evolved earlier than the separation of the domains. Translational initiation, on the other hand, is not shared among the domains. Therefore, it is likely that genes involved in translation had another shared mechanism of initiation prior to the evolution of the AUG-based system, and that they may continue to maintain it by using alternate start codons.

Indeed, there may actually be a need to maintain this earlier system of gene regulation. By observing the tightly-regulated expression of all members of the ribosome, Nomura argued that those genes were regulated through negative feedback at the translational level [31]. Especially under environmental starvation conditions, cells need to reduce translation to survive. It is possible that non-AUG start codons play a role to maintain this feedback control. Rather than opting for the less-restrained AUG-based system, these translationally-related genes may have preferred to maintain their ancestral system of initiation in order to maximize their ability to be regulated.

Many researchers are trying to reconstruct minimal genomes in bacteria in different ways [32–34], primarily to understand the evolution of the universal genetic code and the translational machinery. If the above model is valid, finding genes maintaining non-AUG start codons may offer another way to identify the ancestral form of the translational apparatus.

The reliability of the results presented here depends entirely on the accuracy of gene prediction in the public databases. Here we highlight various sources of errors and biases in our primary data and the measures taken to mitigate their impacts. First, the distribution of assembled bacterial genomes in the public database is skewed heavily toward medically relevant species (*E. coli*, *Salmonella*, *C. difficile*, etc.). At the phylum level, gammaproteobacteria dominate the list for the same reason. Second, the quality of annotations varies between genuses. Annotations in the well-funded and well-studied bacteria like *E. coli* and *B. subtilis* are backed by extensive empirical data, whereas many other rarely studied species rely entirely on computational gene predictions. The prediction of translational start-sites is notoriously error prone [21], compelling other researchers to perform start codon analysis on a much smaller set of curated genomes [29]. Finally, the inconsistent formatting of gene names in various gff files contributes to the potential inaccuracies.

This analysis can be extended to many more microbial species given that there are currently over 250,000 genomes in the NCBI database. However, a number of technical challenges need to be addressed, or otherwise additional genomes will contribute more to noise than signal. These challenges fall into two categories: errors and biases. The errors are contributed by incorrect or incomplete assemblies (10% complete, 40% scaffold, and 50% contig level) and poor annotation quality for incomplete assemblies. Even the labeling of the organisms themselves may not be correct in all cases [35]. The biases appear from the distribution of available organisms toward medically relevant species.

## Methods

A total of 210,130 publicly available prokaryotic genomes were downloaded from the NCBI database in FASTA format along with their corresponding GFF annotations. After excluding all genomes with incomplete assemblies or annotations, 11,904 genomes remained. All gene occurrences from these genomes were extracted based on their corresponding annotations. Genes either missing the gene name attribute or tagged with the pseudo attribute were removed. Only the first feature was considered for genes with multiple consecutive CDS features. This analysis produced a total of 3,818,377 gene occurrences across 11,904 genomes for which the start codon nucleotide sequences were extracted. For each gene occurrence, its gene name, start codon, start position, end position, and source file name were captured in a separate table.

To break down the start codon frequencies phylogenetically, we ultimately concluded that it would be most insightful to categorize gene-start codon pairs based on the genus from which they were sequenced. For each gene within a genus, a normalized value ("non-AUG ratio") was introduced, which measured the ratio of its number of non-AUG start codon occurrences to its total number of occurrences. Using this normalized value reduced biases from those genuses with a high number of sequenced genomes. Furthermore, to avoid bias from phyla with many sequenced genuses, this analysis was limited to the 30 genuses selected from proteobacteria, firmicutes, actinobacteria, planctomycetes and tenericutes. Only those genuses with high total gene occurrences were selected and this minimized noise from the genuses with few prior annotations.

The final 36 non-AUG genes are those who were present in at least two-thirds of the 30 genuses (at least 20), and who had a non-AUG ratio of more than 0.5 in at least a third of the 30 genuses (at least 10).

All code and data used in this work are publicly available from gitlab (https://gitlab.com/anne__g/start-codon-analysis).

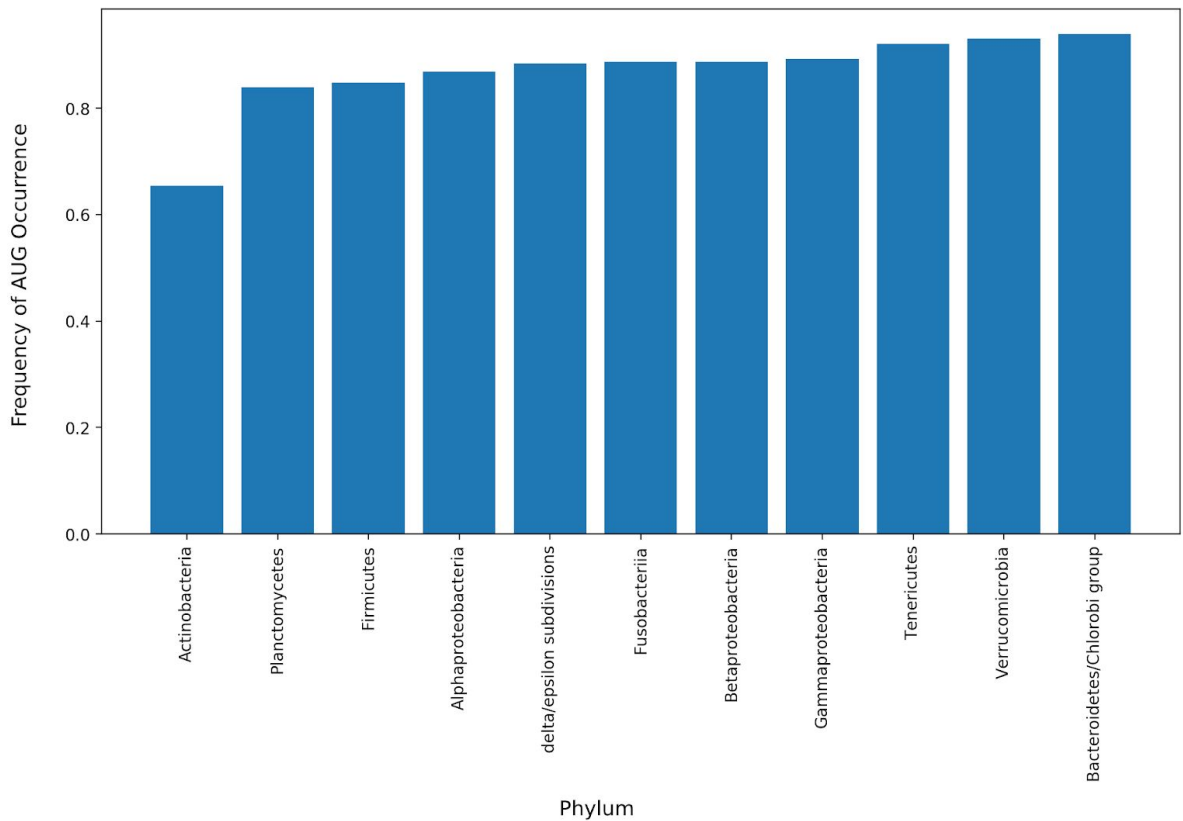

**Figure 1.** A graphical representation of the preliminary distribution of start codon frequencies across bacterial genomes, categorized based on the phylum to which the genome belonged. The bars represent the ratio of the total number of genes using AUG start codons to the total number of genes within a phylum.

| Superphylum | Phylum | Genus |
|---|---|---|
| PVC Group | Planctomycetes | Planctomycetes |
| Terrabacteria | Tenericutes | Mycoplasma |
| Terrabacteria | Actinobacteria | Corynebacterium, Mycobacterium, Streptomyces, Bifidobacterium |
| Terrabacteria | Firmicutes | Bacillus, Lactobacillus, Listeria, Staphylococcus, Streptococcus, Paenibacillus, Clostridium, Brevibacillus, Clostridioides, Enterococcus |

| Proteobacteria | Betaproteobacteria | Bordetella, Neisseria |
| --- | --- | --- |
| Proteobacteria | Delta/Epsilon Subdivisions | Campylobacter, Helicobacter |
| Proteobacteria | Gammaproteobacteria | Mannheimia, Pasteurella, Serratia, Citrobacter, Enterobacter, Escherichia, Klebsiella, Salmonella, Xanthomonas, Pseudomonas |

**Table 1.** The list of the 30 genuses we considered in our research, organized by their superphylum and phylum.

| Gene | Detailed Function | Functional Category | Effect of Deletion in *E. coli* |
| --- | --- | --- | --- |
| infC | translation initiation factor | translation | essential |
| rpoC | RNA polymerase subunit beta | transcription | essential |
| rnpA | RNase P protein component | translation | essential |
| hisS | Histidine-tRNA ligase | translation | essential |
| rimP | ribosome maturation factor | translation | reduced growth at high-temperature |
| recJ | ssDNA-specific exonuclease | replication | viable |
| atpF | ATP synthesis | other | viable |
| bipA (typA) | 50S ribosomal subunit assembly factor | translation | reduced growth |
| dnaA | Chromosomal replication | replication | essential |
| rpsT | 30S ribosomal subunit | translation | essential |
| rsfS | ribosomal silencing factor | translation | substantially reduced cell viability |
| uvrC | DNA damage repair | replication | viable |

| hflX | ribosome rescue factor | translation | viable |
| --- | --- | --- | --- |
| crcB | F(-) channel | other | essential |
| rsgA | ribosome small subunit-dependent GTPase A | translation | essential |
| carA | carbamoyl phosphate synthetase subunit alpha | other | viable |
| infB | translation initiation factor IF-2t | translation | essential |
| rlmH | 23S rRNA m(3)psi1915 methyltransferase | translation | viable |
| argS | arginine-tRNA ligase | translation | essential |
| pnp | polynucleotide phosphorylase | transcription | reduced growth |
| pxpB | 5-oxoprolinase component B | other | slows growth on minimal medium |
| tuf | translation elongation factor | translation | viable |
| secF | Sec translocon accessory complex subunit | protein export | disruption of secD and secF confers cold-sensitive growth |
| rpoB | RNA polymerase subunit beta | transcription | essential |
| secD | Sec translocon accessory complex subunit | protein export | disruption of secD and secF confers cold-sensitive growth |
| fdhD | sulfurtransferase for molybdenum cofactor sulfuration | other | exhibit defect in FDH-H activity |
| rsmH | 16S rRNA m(4)C1402 methyltransferase | translation | viable |
| rsmG | 16S rRNA m(7)G527 methyltransferase | translation | low level streptomycin resistance |
| murE | UDP-N-acetylmuramoyl-L-alanyl-D-glutamate--2%2C6-diaminopimelate ligase | other | essential |
| rpsN | 30S ribosomal subunit | translation | essential |
| nusB | transcription antitermination protein | transcription | essential |

| grpE | nucleotide exchange factor | protein folding | essential |
|------|---------------------------|-----------------|-----------|
| ispG | (E)-4-hydroxy-3-methylbut-2-enyl-diphosphate synthase (flavodoxin) | other | essential |
| radA | DNA recombination protein | replication | reduced survival after ionising irradiation |
| plsX | putative phosphate acyltransferase | other | viable |
| rpsC | 30S ribosomal subunit | translation | essential |

**Table 2.** The top 36 "non-AUG genes" are presented with a) their corresponding functions, b) information on whether or not they are functionally related to translation or DNA replication, and c) whether they are identified as essential in *E. coli*.

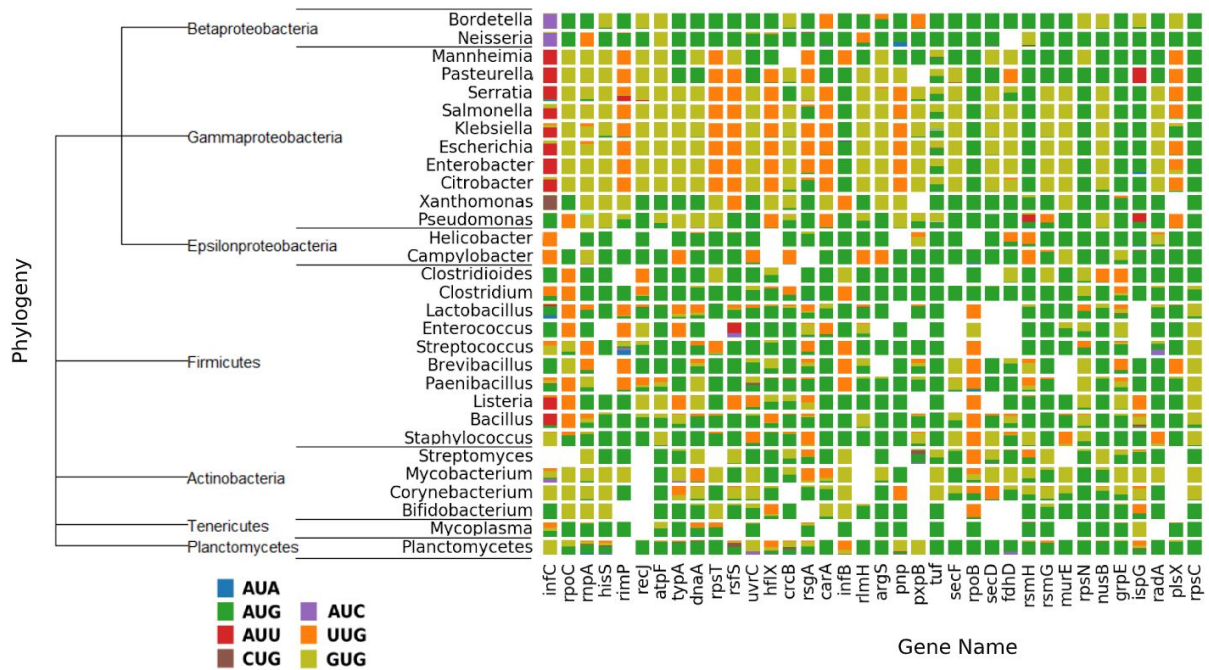

**Figure 2.** A table comparing the start codon frequencies for each gene-genus combination. Genuses are organized phylogenetically; the top 36 non-AUG genes are ordered according to their tendency to use a non-standard start codon. Blank squares occurred where the gene was not present in the genus among the selected gff files.